\documentclass[aps,reprint,longbibliography, superscriptaddress]{revtex4-1}
\usepackage[T1]{fontenc}
\usepackage[latin9]{inputenc}
\usepackage{units}
\usepackage{textcomp}
\usepackage{amsmath}
\usepackage{amssymb}
\usepackage{graphicx}
\usepackage{wasysym}
\usepackage[dvipsnames]{xcolor} 
\usepackage[unicode=true,pdfusetitle,
 bookmarks=true,bookmarksnumbered=false,bookmarksopen=false,
 breaklinks=true,pdfborder={0 0 0},pdfborderstyle={},backref=false,colorlinks=true,
 anchorcolor=black, 
 linkcolor=BrickRed,  
 citecolor=NavyBlue,  
 urlcolor=Blue]  
 {hyperref}
 
 \AtBeginDocument{
    \newwrite\bibnotes
    \def\bibnotesext{Notes.bib}
    \immediate\openout\bibnotes=\jobname\bibnotesext
    \immediate\write\bibnotes{@CONTROL{REVTEX41Control}}
    \immediate\write\bibnotes{@CONTROL{
    apsrev41Control,author="08",editor="1",pages="1",title="0",year="1"}}
     \if@filesw
     \immediate\write\@auxout{\string\citation{apsrev41Control}}
    \fi
}



\begin{document}

\title{Frictional Effects on RNA Folding: Speed Limit and Kramers Turnover}

\author{Naoto Hori}
\email{hori.naoto@gmail.com}
\affiliation{Department of Chemistry, University of Texas, Austin, Texas 78712, United States}

\author{Natalia A. Denesyuk}
\affiliation{Biophysics Program, Institute for Physical Science and Technology, University of Maryland, College Park, Maryland 20742, United States} 

\author{D. Thirumalai}
\email{dave.thirumalai@gmail.com}
\affiliation{Department of Chemistry, University of Texas, Austin, Texas 78712, United States}

\date{September 1, 2018}

\begin{abstract}
We investigated frictional effects on the folding rates of a human telomerase hairpin (hTR HP) and H-type pseudoknot from the Beet Western Yellow Virus (BWYV PK) using simulations of the Three Interaction Site (TIS) model for RNA. The heat capacity from TIS model simulations, calculated using temperature replica exchange simulations, reproduces nearly quantitatively the available experimental 
data for the hTR HP. 
The corresponding results for BWYV PK serve as predictions. We calculated the folding rates ($k_\mathrm{F}$) from more than 100 folding trajectories for each value of the solvent viscosity ($\eta$) at a fixed salt concentration of 200 mM. By using the theoretical estimate ($\propto$$\sqrt{N}$ where $N$ is the number of nucleotides) for folding free energy barrier, $k_\mathrm{F}$ data for both the RNAs are quantitatively fit using one-dimensional Kramers' theory with two parameters specifying the curvatures in the unfolded basin and the barrier top. In the high-friction regime ($\eta\gtrsim10^{-5}\,\textrm{Pa\ensuremath{\cdot}s}$),
for both HP and PK, $k_\mathrm{F}$s decrease  as $\nicefrac{1}{\eta}$ 
whereas in the low friction regime, $k_\mathrm{F}$ values increase as $\eta$ increases, leading to a maximum folding rate
at a moderate viscosity ($\sim10^{-6}\,\textrm{Pa\ensuremath{\cdot}s}$), 
which is the Kramers turnover. 
From the fits, we find that the speed limit to RNA folding at water viscosity is between 1 and 4 $\mathrm{\mu s}$, which is in accord with our previous theoretical prediction as well as results from several single molecule experiments. Both the RNA constructs fold by parallel pathways.  Surprisingly, we find that the flux through the 
pathways could be altered by changing solvent viscosity, a prediction that is more easily testable in RNA than in proteins. 

\begin{center}
\includegraphics{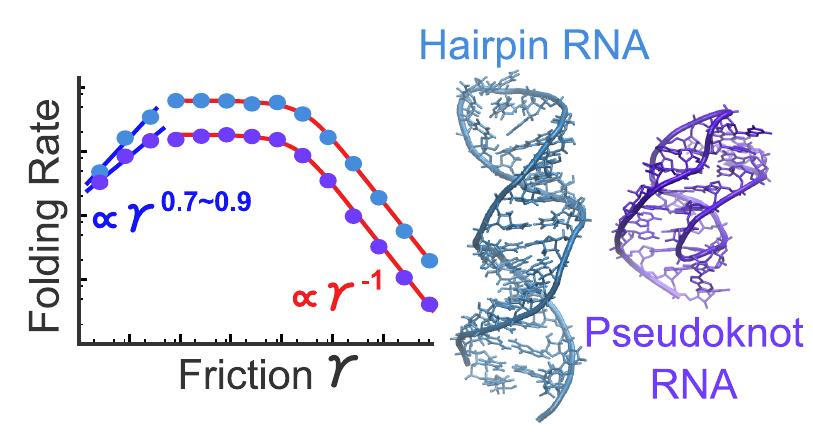}
\end{center}
\end{abstract}

\maketitle


\section*{Introduction}

The effects of friction on barrier crossing events, with a rich
history \cite{Berne1988,Hanggi90RMP}, have also been used to obtain
insights into the dynamics and folding of proteins. For example, in
a pioneering study, Eaton and co-workers established that accounting
for the internal friction is needed to explain experiments in the
ligand recombination to the heme in myoglobin \cite{Ansari1992S}.
Only much later, the importance of internal friction, a concept introduced
in the context of polymer physics \cite{DeGennes79}, in controlling
the dynamics of folded and unfolded states of proteins has been appreciated
in a number of experimental \cite{Soranno12PNAS,Chun13Nature,Schuler16ARB,Soranno17PNAS}
and theoretical \cite{Sagnella00JCP,Echeverria14JACS,Zheng15JACS,Avdoshenko17SciRep,Daldrop18PNAS}
studies. 
The presence of internal friction is typically identified as a deviation 
in the viscosity ($\eta$) dependence of reaction rates from the predictions based on Kramers' theory \cite{Kramers40Physica}.
The timeless Kramers' theory showed that the rate should increase
linearly with $\eta$ at small $\eta$ and decrease as $\nicefrac{1}{\eta}$
at large $\eta$. The change from small $\eta$ behavior to $\nicefrac{1}{\eta}$
dependence with a maximum at intermediate viscosity values is often
referred to as the Kramers turnover \cite{Berne1988,Hanggi90RMP,Grote80JCP}.
Theoretical studies \cite{Klimov1997,Plotkin98PRL} also showed
that folding rates of the so-called two-state folders are in accord
with the theory of Kramers \cite{Kramers40Physica}.

Kramers' theory has been used to understand frictional effects of
the solvent in various reactions, from diffusion of single particles
to folding of proteins that are more complex with the multidimensional
folding landscape. Although Kramers' theory was originally developed
for barrier crossing in a one-dimensional potential with a single
barrier, experiments and simulations suggest the theory holds for
dynamic processes in biomolecules. Interestingly, following the theoretical
study,  establishing that folding rates ($k_\mathrm{F}$) of proteins vary
as $k_\mathrm{F}\sim\nicefrac{1}{\eta}$ \cite{Klimov1997}, experiments on
cold shock protein \cite{Jacob99NSB}, chymotrypsin inhibitor \cite{Ladurner99NSB},
and protein L \cite{Plaxco99NSB} confirmed Kramers' high-$\eta$ predictions.
Although these studies showed that the rate dependence on $\eta$
follows Kramers' prediction, this was most vividly demonstrated in
single molecule studies only recently by Chung and Eaton \cite{Chun13Nature}.
The success of the Kramers' theory, which views the complex process
of polypeptide chain organization as diffusion in an effective one-dimensional 
landscape, is surprising. However, it has been shown using
lattice models \cite{Socci96JCP} that diffusion in an energy landscape
as a function of a collective coordinates, such as the fraction of
native contact ($Q$), provides an accurate description of the folding
rates obtained in simulations.  Subsequently, computational studies \cite{Best2006}
using G\={o} model for a helix bundle further showed that the rate
dependence follows the theoretical predictions including the Kramers
turnover, providing additional justification that $Q$ is a good reaction
coordinate for protein sequences that are well optimized. 

In contrast to several studies probing viscosity effects on protein
folding and dynamics, frictional effects on nucleic acid folding have
been much less studied. 
A vexing issue in experiments is that common viscogens such as 
glycerol may significantly alter the stability of RNA molecules. 
Thus, in order to isolate the frictional effects, a condition of isostability has to be established 
by manipulating other experimental parameters such as temperature 
to compensate for the stability change caused by adding viscogens \cite{Ansari05JPCB}.
Ansari and Kuznetsov showed that, when corrected
for stability changes, the rates of hairpin formation of a DNA sequence
are proportional to $\nicefrac{1}{\eta}$ \cite{Ansari05JPCB}.
Kramers' predictions at high $\eta$ were also borne out in the folding
of G-quadruplex DNA \cite{Lannan2012}, and most recently in a tetraloop-receptor
formation in RNA \cite{Dupuis2018JPCB}.
These studies show that nucleic acid folding might
also be viewed as diffusion in an effective one-dimensional folding
landscape.

In this paper, we consider frictional effects on RNA folding using
coarse-grained (CG) simulations. We investigate the variations in
rates of folding of a human telomerase hairpin (hTR HP) and
an H-type pseudoknot from beet western yellow virus (BWYV
PK) as a function of $\eta$. Because both the HP and PK fold by parallel
pathways, our study allows us to examine whether frictional effects
affect the flux through parallel pathways in RNA folding. Despite
the differences in sequences and the folded structures, the dependence
of $k_\mathrm{F}$ on $\eta$ is quantitatively fit using Kramers' theory
including the predicted turnover. The excellent agreement between
theory and simulations allows us to estimate a speed limit for RNA,
which we find to be $1\sim4\,\mathrm{\mu s}$. Surprisingly, we find
that the flux through the pathways may be altered by changing solvent
viscosity for both the HP and PK. The change in the flux is more pronounced
for HP, especially at a temperature below the melting temperature.
We argue that this prediction is amenable to experimental tests in
RNA even though it has been difficult to demonstrate it for protein
folding.

\section*{Materials and Methods}

\begin{figure}
\includegraphics[width=\columnwidth]{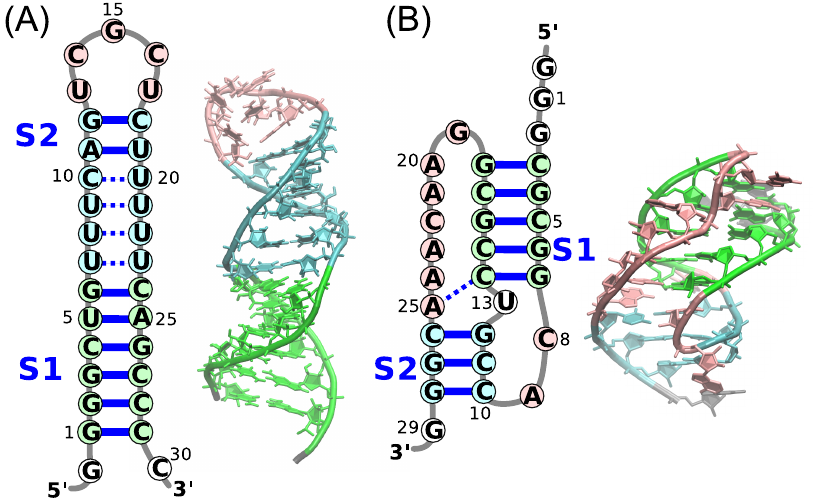}
\caption{\label{fig:structure}\textbf{(A)} Secondary representation and sequence of
human telomerase hairpin (hTR HP). The folded hairpin structure is on the right. 
Note that there are four noncanonical base pairs between S1 and S2.
\textbf{(B)} Secondary structure of Beet Western Yellow Virus pseudoknot (BWYV PK). 
The tertiary structure of the PK is shown on the right. 
In the secondary structures, blue lines represent canonical base pairs (thick lines) 
and noncanonical pairs (dotted lines).}
\end{figure}

\paragraph*{\textbf{RNA Molecules:}}
We choose a sequence that forms a hairpin (HP) with no tertiary interactions
from the human telomerase (hTR) and an H-type BWYV pseudoknot (PK),
which is a minimal RNA motif with tertiary interactions. The folding
mechanisms of PKs are diverse \cite{Cho2009}, and they often reach
the native structure by parallel pathways \cite{Roca2018PNAS}. The
use of two RNA molecules with different folded states, with both HP
and PK folding occurring by parallel pathways, allows us to examine
many consequences of viscosity effects on their folding. The structure
of hTR HP (PDB ID 1NA2) has been determined using NMR (see Figure
\ref{fig:structure}A) \cite{Theimer2003PNAS}. The folded structure
of the BWYV PK is taken from the crystal structure (PDB ID 437D) \cite{Su1999}.
The PK has 28 nucleotides forming two stems. The two loop regions
have hydrogen bonding interactions with the stems (Figure~\ref{fig:structure}B).
We added an additional guanosine monophosphate to both the $5^{\prime}$
and $3^{\prime}$ terminus to minimize end effects. Thus, the simulated
PK has 30 nucleotides.

\paragraph*{\textbf{Three Interaction Site (TIS) Model for RNA:}}
We employed a variant of the TIS model, which has been previously
used to make several quantitative predictions for RNA molecules ranging
from hairpins to ribozymes \cite{Hyeon2005,Cho2009,Biyun2011,Denesyuk2015}.
We incorporated the consequences of counterion condensation into
the TIS model, allowing us to predict the thermodynamic properties
of RNA hairpins and PKs that are in remarkable agreement with experiments \cite{Denesyuk2013}.
Because the details of the model have been reported previously, we
only provide a brief description here. In the TIS model \cite{Hyeon2005},
each nucleotide is represented by three coarse-grained spherical beads
corresponding to phosphate (P), ribose sugar (S), and a base (B).
Briefly, the effective potential energy (for details see Ref.~\cite{Denesyuk2013})
of a given RNA conformation is $U_{\textrm{TIS}}=U_{\textrm{L}}+U_{\textrm{EV}}+U_{\textrm{ST}}+U_{\textrm{HB}}+U_{\textrm{EL}}$,
where $U_{\textrm{L}}$ accounts for chain connectivity and angular
rotation of the polynucleic acids, $U_{\textrm{EV}}$ accounts for
excluded volume interactions of each chemical group, and $U_{\textrm{ST}}$
and $U_{\textrm{HB}}$ are the base-stacking and hydrogen-bond interactions,
respectively.

Electrostatic interactions between the phosphate (P) groups are given
by $U_{\textrm{EL}}$. The repulsive electrostatic interactions between
the P sites are taken into account through the Debye-H\"{u}ckel theory,
$U_{\textrm{EL}}=\sum_{i,j}\frac{q^{\ast2}e^{2}}{4\pi\varepsilon_{0}\varepsilon(T)r_{ij}}\exp\left(-\frac{r_{ij}}{\lambda_{\textrm{D}}}\right)$,
where the Debye length is $\lambda_{\textrm{D}}=\sqrt{\frac{\varepsilon_0\varepsilon(T)k_{\textrm{B}}T}{2e^{2}N_{\mathrm{A}}I}}$.
In the present simulations, salt concentration (monovalent ions) is
set to 200 mM, which is close to the physiological value. 
The ionic strength $I=\frac{1}{2}\sum c_{i}z_{i}^{2}$ where $c_{i}$
is the molar concentration, and $z_{i}$ is the charge number of ion
$i$, and the sum is taken over all ion types. Following our earlier
study \cite{Denesyuk2013}, we used an experimentally fit function
for the temperature-dependent dielectric constant $\varepsilon(T)$ \cite{Malmberg1956}.
To account for counterion condensation, we used a renormalized charge
on the phosphate group, $-q^{\ast}e\,(q^{\ast}<1)$. The renormalized
value of the charge on the P group is approximately given by $-q^{\ast}(T)e=\frac{-be}{l_{\textrm{B}}(T)}$,
where the Bjerrum length is $l_{\textrm{B}}(T)=\frac{e^{2}}{4\pi\varepsilon_0\varepsilon(T)k_{\textrm{B}}T}$,
and $b$ is the mean distance between the charges on the phosphate
groups \cite{Manning1969}. We showed elsewhere \cite{Denesyuk2013}
that a constant value of $b=0.44$ nm accounts for the thermodynamics
of several RNA molecules, and is the value adopted here. All the force-field
parameters used here are the same as in our earlier study \cite{Denesyuk2013}.

\paragraph*{\textbf{Simulation Details:}}
We performed Langevin dynamics simulations by solving the equation
of motion, 
\begin{equation}
m\ddot{\boldsymbol{x}}=-\frac{\partial U_{\textrm{TIS}}}{\partial\boldsymbol{x}}-\gamma\dot{\boldsymbol{x}}+\boldsymbol{\Gamma},\label{eq:Langevin}
\end{equation}
where $m$ is the mass of the particle, $\boldsymbol{x}$  is the
coordinate, and $\boldsymbol{\Gamma}$ is a Gaussian random force
that satisfies the fluctuation-dissipation relation given by $\left\langle \boldsymbol{\Gamma}_{i}(t)\boldsymbol{\Gamma}_{j}(t')\right\rangle =6\gamma k_{\mathrm{B}}T\delta(t-t')\delta_{ij}$.
The friction coefficient follows the Stokes-Einstein relation, $\gamma=6\pi\eta R$,
where $R$ is the appropriate size of the coarse-grained bead (P,
S and B) and $\eta$ is the solvent viscosity. The numerical integration
is performed using the velocity-Verlet algorithm \cite{Honeycutt1992}.

In the high friction regime where $\eta=10^{-3}$ to $10^{-2}\,\mathrm{Pa\cdot s}$, we performed
Brownian dynamics simulations \cite{Ermak1978} by neglecting the inertial
term, since the dynamics is overdamped. In this limit, the equation
of motion is, 
\begin{equation}
\dot{\boldsymbol{x}}=-\frac{1}{\gamma}\frac{\partial U_{\textrm{TIS}}}{\partial\boldsymbol{x}}+\boldsymbol{\Gamma}.\label{eq:Brownian}
\end{equation}

We used reduced units
in the analysis of data \cite{Klimov1997}. In the TIS representation, we chose the
mass of a bead $m=$ 116 g/mol, the typical length scale $a=0.4$
nm, and the energy scale $\varepsilon=1$ kcal/mol. Thus,
the natural measure for time in Eq.~\ref{eq:Langevin} is $\tau=(ma^{2})^{\nicefrac{1}{2}}\sim2\,\mathrm{ps}$.
In the overdamped condition (Eq.~\ref{eq:Brownian}), the natural unit of time is $\tau=\frac{\gamma a^{2}}{k_{B}T}$.
We used this measure to obtain $\tau\approx300\,\mathrm{ps}$ for converting the simulation times to real times
at the viscosity of water, $\eta_{w}=10^{-3}\,\mathrm{Pa\cdot s}$ \cite{Hyeon08JACS}. 

 We confirmed that both Langevin dynamics and Brownian dynamics simulations
give identical results at $\eta=10^{-3}\,\textrm{Pa\ensuremath{\cdot}s}$,
using simulations of hTR hairpin. The difference between the two simulations
method is in the range of statistical error estimated by the jack-knife
method. For example, the folding rate for hTR HP is $k_{\textrm{F}}=5.5\pm0.5\,\textrm{ms}^{-1}$
calculated from 100 trajectories generated using Brownian dynamics simulations 
and is $k_{\textrm{F}}=6.5\pm0.6\,\textrm{ms}^{-1}$ obtained from another
set of 100 trajectories generated using Langevin dynamics simulations
at $\eta=10^{-3}\,\textrm{Pa\ensuremath{\cdot}s}$. 

\paragraph*{\textbf{Hydrodynamic Interactions:}}
In order to ensure that the results are robust, we did limited simulations
of folding by including hydrodynamic interactions (HI). To take into
account the effects of HI, we performed Brownian dynamics simulations
using the following form with conformation-dependent mobility tensor,
\begin{equation}
\dot{\boldsymbol{x}_{i}}=-\sum_{j}\boldsymbol{\mu}_{ij}\frac{\partial U_{\textrm{TIS}}}{\partial\boldsymbol{x}_{i}}+\boldsymbol{\Gamma},\label{eq:HI}
\end{equation}
where $\boldsymbol{\mu}$, the mobility tensor, is computed using
the Rotne-Prager-Yamakawa approximation \cite{Ermak1978},
\begin{equation}
\boldsymbol{\mu}_{ij}=\begin{cases}
\frac{1}{6\pi\eta R} & (i=j)\\
\frac{1}{8\pi\eta r_{ij}}\left[\left(\boldsymbol{1}+\frac{\boldsymbol{r}_{ij}\boldsymbol{r}_{ij}}{r_{ij}^{2}}\right)+\frac{2R^{2}}{r_{ij}^{2}}\left(\frac{\boldsymbol{1}}{3}-\frac{\boldsymbol{r}_{ij}\boldsymbol{r}_{ij}}{r_{ij}^{2}}\right)\right] & (i\neq j)
\end{cases}.\label{eq:HI_mu}
\end{equation}
In the above equation, $\boldsymbol{r}_{ij}$ is a coordinate vector
between beads $i$ and $j.$ Because coarse-grained beads in our TIS
model have different radii ($R$) depending on the type of beads (phosphate,
sugar, and bases) \cite{Denesyuk2013}, we employed a modified form
of $\boldsymbol{\mu}$ developed by Zuk \textit{et al.} \cite{Zuk2014}.

\paragraph*{\textbf{Thermodynamics Properties:}}
We performed temperature-replica-exchange simulations (T-REMD) \cite{Sugita1999}
to calculate the heat capacity. Temperature was distributed from 0
to 120 \textdegree C with 16 replicas at 200 mM salt concentration.
The T-REMD simulation is performed using a lower friction ($\eta=10^{-5}\,\textrm{Pa\ensuremath{\cdot}s}$)
to enhance the efficiency of conformational sampling \cite{Honeycutt1992}.

\paragraph*{\textbf{Order Parameter:}}
In order to determine if a folding reaction is completed, we used
the structural overlap function \cite{Camacho93PNAS} 
\begin{equation}
\chi=\frac{1}{N_{p}}\sum_{i,j}^{N_{p}}H\left(d-\left|r_{ij}-r_{ij}^{0}\right|\right),\label{eq:SO}
\end{equation}
 where $H$ is the Heaviside step function, $d=0.25$ nm is the tolerance,
and $r_{ij}^{0}$ is the distance between particles $i$ and $j$
in the native structure. The summation is taken over all pairs of
coarse-grained sites separated by two or more covalent bonds, and
$N_{p}$ is the number of such pairs. The structural overlap function
quantifies the similarity of a given conformations to the native conformation.
It is unity if the conformation is identical to the native state.
In $T$-quench kinetics simulations, if the value of the structural overlap
function exceeds a threshold, $\left\langle \chi\right\rangle _{T_\mathrm{L}}$,
the trajectory is deemed to be completed, and the folding time $\tau_{i}$
is recorded; $\left\langle \chi\right\rangle _{T_\mathrm{L}}$ is the thermodynamic
average at the lower simulation temperature at which RNA molecules
are predominantly folded (Table~\ref{tab:T}). In addition to $\chi$,
we also calculated the average value of $\left\langle R_{\mathrm{g}}\right\rangle $,
measurable in scattering experiments (SAXS or SANS), to assess the temperature dependence
of compaction of the RNA molecules. 

\begingroup
\setlength{\tabcolsep}{5pt}
\renewcommand{\arraystretch}{1.5} 

\begin{table}
\caption{\label{tab:T}Thermodynamic Properties of the RNAs. }
\begin{tabular}{rcccccc}
\hline 
 & $N$ & $T_\mathrm{L}$  & $T_\mathrm{m1}$  & $T_\mathrm{m2}$  $^{a}$ & $\left\langle \chi\right\rangle _{T_\mathrm{L}}$ & $\Delta G^{\ddagger}$ $^{ab}$ \tabularnewline
\hline 
hTR HP & 31 & 22  & 55 (54)$^{c}$ & 81 (74)$^{c}$ & 0.63 & 5.1 \tabularnewline
BWYV PK & 30 & 20  & 52 & 90 & 0.74 & 5.0 \tabularnewline
\hline 
\end{tabular}

\begin{raggedright}
	$^{a}$ The temperatures are in the units of $^{\circ}\mathrm{C}$ and $\Delta G^{\ddagger}$ in $k_{\mathrm{B}}T$.
	$^{b}$ The free energy barrier $\Delta G^{\ddagger}$ is estimated based on the number of nucleotides, $N$ \cite{Hyeon2012BJ}.
	$^{c}$ For hTR hairpin, melting temperatures measured in experiments \cite{Theimer2003PNAS} are shown in parentheses.
\par
\end{raggedright}

\end{table}
\endgroup

\paragraph*{\textbf{$T$-Quench Folding:}}
To prepare the initial structural ensemble for $T$-quench simulations,
we first performed low-friction Langevin dynamics simulations. The
simulation temperatures are chosen to be 1.2 times higher than the
second melting temperature (in Kelvin unit) to ensure that completely
unfolded conformations are populated. After generating a sufficiently
long trajectory to ensure that the chain has equilibrated, the unfolded
conformations are sampled every $10^{5}$ time steps. Finally, we
collected hundreds of conformations, which were used as initial structures
in the $T$-quench folding simulations.

In order to initiate folding, starting from an unfolded structure,
we quenched the temperature to $T_{\textrm{S}}$ and generated folding
trajectories using Langevin or Brownian dynamics simulations by varying
the solvent viscosity from $\eta=3.2\times10^{-9}$ to $10^{-2}$
$\textrm{Pa\ensuremath{\cdot}s}$ (cf. water viscosity $\eta_{\mathrm{w}}\approx10^{-3}$
$\textrm{Pa\ensuremath{\cdot}s}$). The viscosity is directly related
to the friction coefficient as $\gamma=6\pi\eta R$ where $R$ is
the radius of coarse-grained beads. For each condition, at least 100 folding
trajectories are generated. Folding time $\tau_{i}$ is measured by
monitoring the overlap function, $\chi$, in each trajectory $i$,
and folding rates were calculated by averaging over $M$ trajectories,
$k_\mathrm{F}=\tau_{\textrm{MFPT}}^{-1}=\left(\frac{1}{M}\sum_{i}^{M}\tau_{i}\right)^{-1}$ \cite{Reimann1999}.
We used two values of $T_\mathrm{S}$. One is $T_\mathrm{S}=T_\mathrm{m1}$, which is the
lower melting temperature in the heat capacity curve (Figure~\ref{fig:Thermo}),
and the other is $T_\mathrm{S}=T_\mathrm{L}<T_\mathrm{m1}$,
which is obtained by multiplying a factor 0.9 to $T_\mathrm{m1}$ in Kelvin unit.
The values of $T_\mathrm{L}$ and $T_\mathrm{m1}$ are listed in Table~\ref{tab:T}. 

\begin{figure}
\includegraphics[width=0.8\columnwidth]{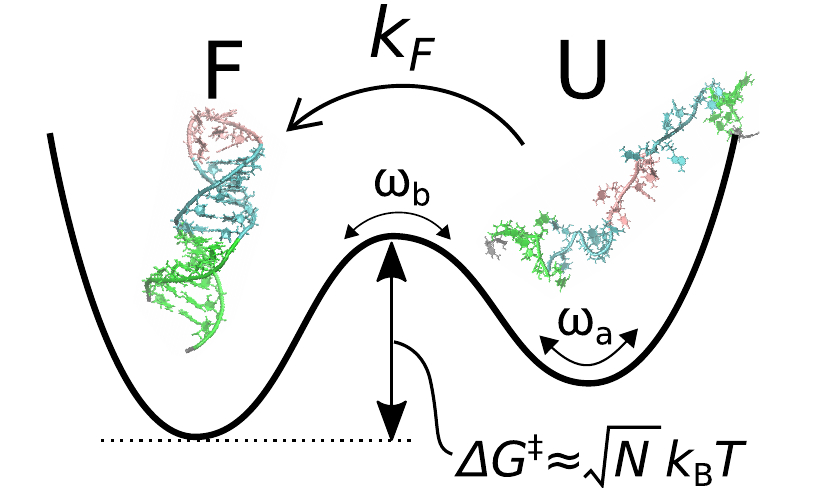}
\caption{\label{fig:schematic1D}Schematic of effective one-dimensional landscape
of RNA folding from the unfolded state (U) to the folded state (F).
Parameters, $\omega_{a}$ and $\omega_{b}$, determine curvatures
of the free energy surface at the reactant basin (U) and the saddle
point. $\Delta G^{\ddagger}$ is the height of the free energy barrier.}
\end{figure}

\paragraph*{\textbf{Data Analysis Using Kramers' Rate Theory:}}
The simulation data for RNA is analyzed using Kramers' theory \cite{Kramers40Physica}
in which  RNA folding is pictured as a barrier crossing event in an
effective one-dimensional landscape (see Figure~\ref{fig:schematic1D}).
For our purposes here, it is not relevant to determine the optimal reaction
coordinate because we estimate the values of the barrier heights theoretically  
and obtain the two frequencies $\omega_a$ and $\omega_b$ (see below for the definition) by fitting the simulation data to the theory.
According to transition state theory (TST), the reaction rate is expressed
as 
\begin{equation}
k_{\textrm{TST}}=\frac{\omega_{a}}{2\pi}\exp\left(\frac{-\Delta G^{\ddagger}}{k_{\mathrm{B}}T}\right),\label{eq:kTST}
\end{equation}
 where $k_{\mathrm{B}}$ is the Boltzmann constant and $T$ is the
temperature. The rate, $k_{\textrm{TST}}$, gives us an upper bound
of the true reaction rate since in the TST there are no
recrossing events once RNA reaches the saddle point \cite{Berne1988,Hanggi90RMP}.

The Kramers' folding rate in the intermediate to high $\eta$ range may be written as
\begin{equation}
k_{\mathrm{KR}}=\frac{\omega_{a}}{2\pi\omega_{b}}\left(\sqrt{\frac{\gamma^{2}}{4}+\omega_{b}^{2}}-\frac{\gamma}{2}\right)\exp\left(\frac{-\Delta G^{\ddagger}}{k_{\mathrm{B}}T}\right),\label{eq:k_KR}
\end{equation}
 where $\gamma$ is the friction coefficient and $\omega_{a}$ and $\omega_{b}$
are the parameters that determine curvatures of the free energy surface
at the reactant basin and saddle point (maximum in the free energy
surface), respectively (Figure~\ref{fig:schematic1D}). It is assumed
that, in the vicinity of the saddle point, the free energy may be approximated
by a parabola $G(x)=G(x_{b})-\frac{1}{2}m\omega_{b}^{2}(x_{b}-x)^{2}$
where $x$ is an unknown reaction coordinate, $x_{b}$ is the position of
the saddle, and  $m\omega_{b}^{2}=-\frac{\partial^{2}G(x)}{\partial x^{2}}$. 

In the high friction limit, 
\begin{equation}
k_{\mathrm{KR}}^{\textrm{H}} \sim \frac{\omega_{a}}{2\pi}\frac{\omega_{b}}{\gamma}\exp\left(\frac{-\Delta G^{\ddagger}}{k_{\textrm{B}}T}\right)\quad\left(\ensuremath{\frac{\gamma}{2}\gg\omega_{b}}\right),\label{eq:kKR_H}
\end{equation}
 which shows that the folding rate should depend on the inverse
of the friction coefficient. When the friction is small, the rate linearly
approaches the TST limit, $k_{\mathrm{KR}}^{\textrm{M}} \sim \frac{\omega_{a}}{2\pi}\exp\left(\frac{-\Delta G^{\ddagger}}{k_{\textrm{B}}T}\right) = k_{\textrm{TST}}\quad(\frac{\gamma}{2}\ll\omega_{b}).$ 

If we further consider the time scale at which local equilibrium is
achieved (very weak damping limit), the rate for the effective one-dimensional landscape becomes \cite{Berne1988}
\begin{equation}
k_{\mathrm{KR}}^{\mathrm{L}} \sim \gamma \frac{\Delta G^{\ddagger}}{k_{\textrm{B}}T} \exp\left(\frac{-\Delta G^{\ddagger}}{k_{\textrm{B}}T}\right).\label{eq:kKR_L}
\end{equation}
In this regime, barrier crossing is controlled by energy diffusion \cite{Zwanzig59PhysFluids},
and the TST is no longer valid. These extremely well-known results,
used to analyze the simulations, can be summarized as follows: Kramers'
theory predicts that $k_{\textrm{KR}}^{\textrm{L}}\propto\gamma$
in the low friction regime, $k_{\textrm{KR}}$ reaches a maximum in
moderate friction ($k_{\mathrm{KR}}^{\mathrm{M}}$), and $k_{\textrm{KR}}^{\textrm{H}}\propto\gamma^{-1}$
in the high friction regime.
The folding rates over the entire range of $\gamma$ can be fit using \cite{Borkovec1985JPC}
\begin{equation}
{k}^{-1}_{\mathrm{F}} = {k_{\mathrm{KR}}^{\mathrm{L}}}^{-1} + {k_{\mathrm{TST}}}^{-1} + {k_{\mathrm{KR}}^{\mathrm{H}}}^{-1}.\label{eq:k_connect}
\end{equation}

\paragraph*{\textbf{Barrier Heights in the Folding Landscape:}}
In order to use Eq.~\ref{eq:k_KR}--\ref{eq:k_connect}  to analyze simulation data, 
the free energy barrier to folding has to be calculated. 
However, estimating barrier heights is nontrivial in complex systems because
the precise reaction coordinate is difficult to calculate or guess, particularly for RNA
in which ion effects play a critical role in the folding reaction.
In order to avoid choosing a specific reaction coordinate, we appeal
to theory to calculate the effective barrier height. One of us has
shown \cite{Thirum95JPI}, which has been confirmed by other studies \cite{Dill11PNAS},
that for proteins the free energy barrier  $\approx\sqrt{N}$ where
$N$ is the number of amino acids. In the context of RNA folding,
we showed that there is a robust relationship between the number of
nucleotides ($N$) and the folding rates, $k_\mathrm{F}\thickapprox k_{0}\exp(-\alpha N^{0.5})$ \cite{Hyeon2012BJ}.
Experimental data of folding rates spanning 7 orders of magnitude
(with $N$ varying from 8 to 414) were well fit to the theory using
$\alpha=0.91$ and $k_{0}^{-1}=0.87\,\textrm{\ensuremath{\mu}s}$
(speed limit for RNA folding) \cite{Hyeon2012BJ}. Therefore, in this
study, we estimate the free energy barrier based on $N$ alone. A
clear advantage of using the theoretical estimate is that it eliminates
the need to devise a reaction coordinate. The values of the barrier
height for the two RNAs are given in Table~\ref{tab:T}. 

\paragraph*{}
In summary, our strategy in this study is to (1) conduct folding simulations using the TIS RNA model to obtain the folding rates by varying the solvent viscosity and then (2) examine the applicability of the Kramers' theory to RNA folding by fitting the rates using Eqs.~\ref{eq:k_KR}~to~\ref{eq:k_connect}. In order to calculate the reaction rates in Kramers' theory, the barrier height $\Delta G^{\ddag}$ and frequencies $\omega_{a}$ and $\omega_{b}$ are needed (Eq.~\ref{eq:k_KR}). This would require an appropriate reaction coordinate to characterize the folding landscape. We eliminate the need for creating a specific reaction coordinate by estimating $\Delta G^{\ddag}$ from the length of RNA $(N)$ based on a previous study \cite{Hyeon2012BJ} and by using the other two quantities $\omega_{a}$ and $\omega_{b}$ as fitting parameters.

\section*{Results }

\begin{figure}
\includegraphics{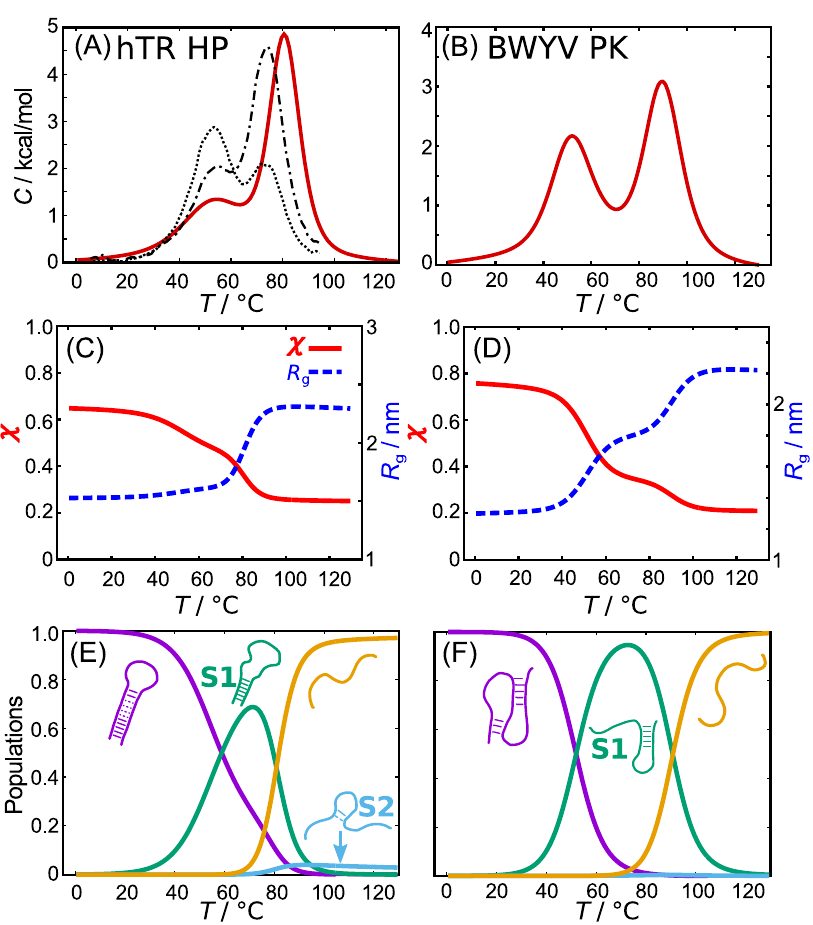}
\caption{\label{fig:Thermo} Temperature dependence of thermodynamic properties
at 200 mM of monovalent salt concentration. \textbf{(A, B)} Heat capacity
of (A) hTR HP and (B) BWYV PK. The red lines are heat capacities,
$C(T)$, computed from the T-REMD simulations. Black lines in panel A 
are UV absorbance melting profiles ($\nicefrac{\delta A}{\delta T}$)
at 260 nm (dotted line) and 280 nm (dot-dash line) experimentally
reported elsewhere \cite{Theimer2003PNAS}. The scale of $\nicefrac{\delta A}{\delta T}$
is not relevant because we only compare the positions of the peaks.
The melting temperatures for the HP are given in Table~\ref{tab:T}.
For the PK, the values of $T_{\textrm{m}}$ are predictions. \textbf{(C, D)}
Structural overlap functions ($\chi$, red solid) and radius of gyration
($R_{g}$, blue dashed). \textbf{(E, F)} Populations of folded (purple),
intermediate (green and cyan), and unfolded states (yellow) as functions
of $T$. }
\end{figure}

\paragraph*{\textbf{Thermal Denaturation:}}
We calculated the heat capacities of the HP and the PK at the monovalent
salt concentration of 200 mM (Figure~\ref{fig:Thermo}). The heat
capacities have two distinct peaks, which indicate there is at least
one intermediate between the unfolded and folded states.
This finding is consistent with previous experimental and simulation
studies \cite{Theimer2003PNAS,Denesyuk2011,Hori2016}. From the position
of the peaks, we determined the two melting temperatures, $T_{\textrm{m1}}$
and $T_{\textrm{m2}}$, whose values are listed in Table~\ref{tab:T}.
It should not go unnoticed that the melting temperatures, $T_\mathrm{m1}$
and $T_\mathrm{m2}$, for the hTR HP are in excellent agreement with experiments,
demonstrating the effectiveness of TIS model in predicting the thermodynamic
properties of RNA. The experimental heat capacity curve is not available
for BWYV PK at 200 mM salt concentration, and hence the values reported
in Table~\ref{tab:T} serve as predictions.

\begin{figure}
\includegraphics{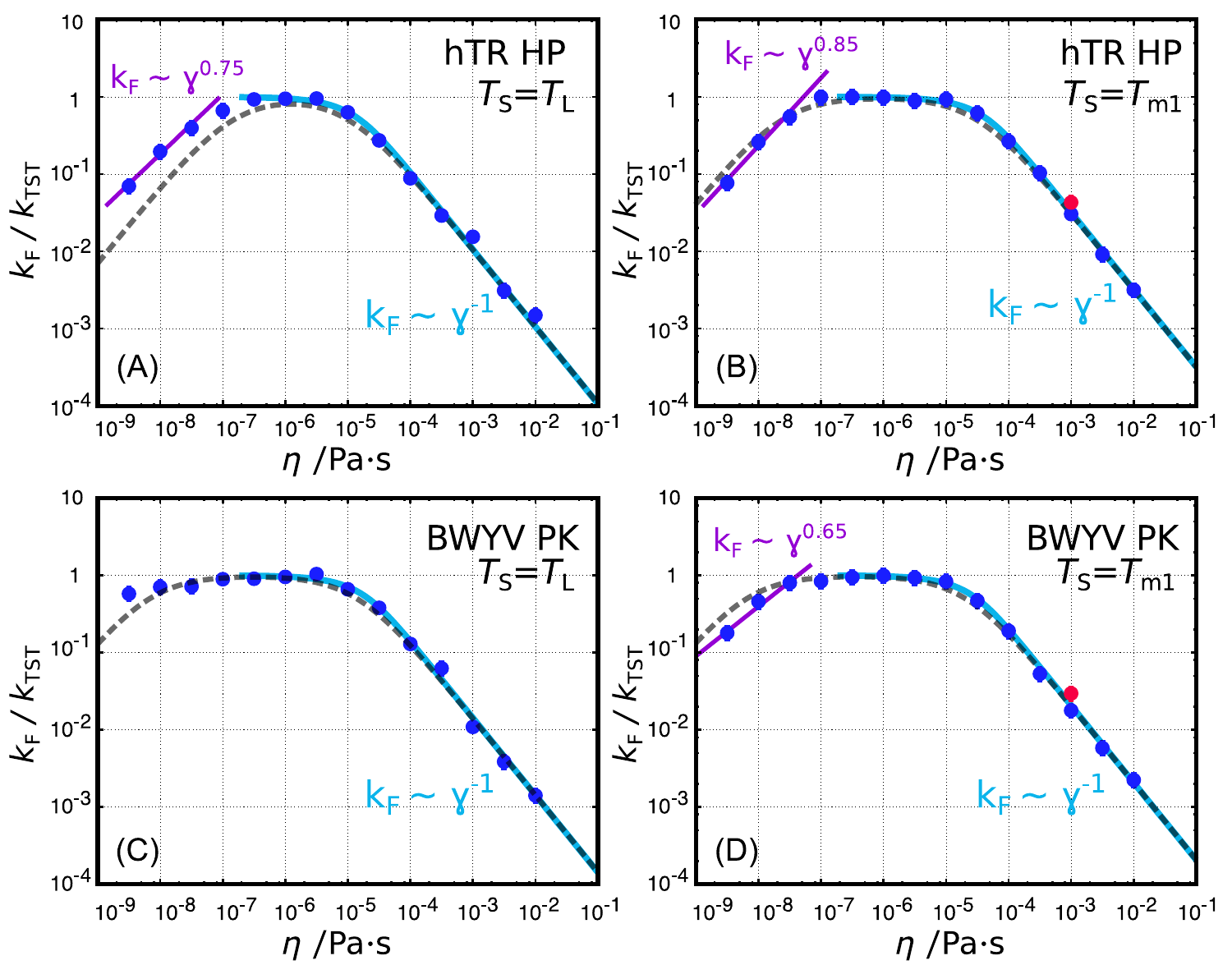}
\caption{\label{fig:kF}
Friction (viscosity) dependence of folding rates at
simulation temperatures $T_\mathrm{S}=T_\mathrm{L}$ and $T_{\textrm{m1}}$ for
hTR HP (A, B) and BWYV PK (C, D). See Table~\ref{tab:T} for
the numerical values of $T_\mathrm{S}$. Folding rates (blue circles) are normalized
by the values from the transition state theory, $k_{\textrm{TST}}$
(Eq.~\ref{eq:kTST}). Error bars, which are presented with 95\% confidence level for each data point, lie within the size of the circles.
The data in the moderate to high friction regime ($\eta\geq10^{-6}\,\textrm{Pa\ensuremath{\cdot}s})$
were fit to Eq.~\ref{eq:k_KR} (lines in cyan). The data in
the low friction regime ($\eta\leq10^{-7.5}\,\mathrm{Pa\cdot s}$)
were fit to Eq.~\ref{eq:kKR_L} except rates of BWYV PK at $T_\mathrm{L}$.
With use of $\omega_{a}$ and $\omega_{b}$ by fitting to Eq.~\ref{eq:k_KR}, 
the rates for the entire range of $\eta$ are well represented by the connecting formula, Eq.~\ref{eq:k_connect} (dashed line).
The results of Brownian dynamics simulations with hydrodynamic interactions
at the water viscosity ($\eta=10^{-3}\,\textrm{Pa\ensuremath{\cdot}s}$)
are shown in red in panels B and D for hTR HP and BWYV PK, respectively.
}
\end{figure}

\paragraph*{\textbf{Viscosity Dependence of the Folding Rates, Kramers Turnover and Absence
of Internal Friction:}}
Friction dependent folding rates obtained from the $T$-quench simulations
are shown in Figure~\ref{fig:kF}. In the high friction regime, $\eta\apprge10^{-5}$
$\textrm{Pa\ensuremath{\cdot}s}$, the folding rates $k_{\textrm{F}}$
decrease as the friction is increased. This behavior is found in both
HP and PK at both temperatures, $T_\mathrm{L}$ and $T_\mathrm{m1}$. In the moderate
friction regime, $10^{-7}\apprle\eta\apprle10^{-5}$ $\textrm{Pa\ensuremath{\cdot}s}$,
the folding rates reach maximum values. For $\eta\geq10^{-6}$ $\textrm{Pa\ensuremath{\cdot}s}$,
we fit the values of $k_{\textrm{F}}$ to Eq.~\ref{eq:k_KR} with
$\Delta G^{\ddagger}\approx0.91\sqrt{N}k_{\mathrm{B}}T$. By adjustment of 
the two free parameters, $\omega_{a}$ and $\omega_{b}$, Eq.~\ref{eq:k_KR}
quantitatively accounts for the simulation data (lines in cyan in
Figure~\ref{fig:kF}, parameters are summarized in Table~\ref{tab:fit}).
Thus, the variation of $k_{\textrm{F}}\propto\eta^{-1}$ in the high
friction regime shows that Kramers' theory accurately describes the
dependence of the folding rates on $\eta$ of these two RNA constructs.
We conclude that even for RNA, driven by electrostatic interactions,
folding could be pictured as a diffusive process in an effective one-dimensional landscape. 
The quantitative account of simulation data on $k_{\textrm{F}}$ using Kramers' theory at high
$\eta$ shows the absence of internal friction in the folding process of these RNA constructs.

\begingroup
\setlength{\tabcolsep}{4pt}
\renewcommand{\arraystretch}{1.2} 
\begin{table}
\caption{\label{tab:fit}Fitting Parameters.}
\begin{tabular}{cc | cccc}
\hline 
 &  & $\omega_{a}$ & $\omega_{b}$ & $\alpha$ & $\frac{\omega_{a}\omega_{b}}{2\pi\gamma}$ at $\eta_{\mathrm{w}}$\tabularnewline
\hline 
hTR HP & $T_\mathrm{L}$ (22$^{\circ}$C) & 0.38 & 0.87 & 0.75 & 3.1 $\mathrm{\mu s^{-1}}$\tabularnewline
 & $T_\mathrm{m1}$ (55$^{\circ}$C) & 0.062 & 2.6 & 0.86 & 1.5 $\mathrm{\mu s^{-1}}$\tabularnewline
\hline 
BWYV PK & $T_\mathrm{L}$ (20$^{\circ}$C) & 0.018 & 1.2 & 0.094 & 0.19 $\mathrm{\mu s^{-1}}$\tabularnewline
 & $T_\mathrm{m1}$ (52$^{\circ}$C) & 0.017 & 1.7 & 0.65 & 0.27 $\mathrm{\mu s^{-1}}$\tabularnewline
 \hline 
\end{tabular} 
\end{table}
\endgroup

\onecolumngrid

\twocolumngrid

As $\eta$ decreases, there is a maximum in
the rate followed by a decrease in $k_{\textrm{F}}$ at low $\eta$,
which shows the expected Kramers turnover (Figure~\ref{fig:kF}).
For $\eta \le 10^{-7}$ $\textrm{Pa\ensuremath{\cdot}s}$, the dependence of the rates is $k_{\textrm{F}}\propto\eta^{\alpha}$
with a positive $\alpha$ (lines in purple in Figure~\ref{fig:kF}, and values of $\alpha$ are in Table~\ref{tab:fit}).
In contrast to the high friction case, the low $\eta$ dependence, that is, $\alpha$ value, varies with each RNA molecule.
When the $\omega_{a}$ and $\omega_{b}$ obtained from the fitting to Eq.~\ref{eq:k_KR} in the high friction regime are used,
the rates for the entire range of $\eta$ are well described by the connection formula, Eq.~\ref{eq:k_connect} (dashed line in Figure~\ref{fig:kF}).
At $T = T_\mathrm{m1}$, the low $\eta$ dependence is in quantitative accord with the theory.
This is remarkable because there is no additional fitting parameter in Eq.~\ref{eq:k_connect} to account for the dependence of $\eta^{\alpha}$ in the low $\eta$ regime.
Although there are some deviations at $T = T_\mathrm{L}$ case, the overall rate dependence showing turnover at moderate friction is well characterized by the Kramers'  theory.  

\begin{figure}
\includegraphics{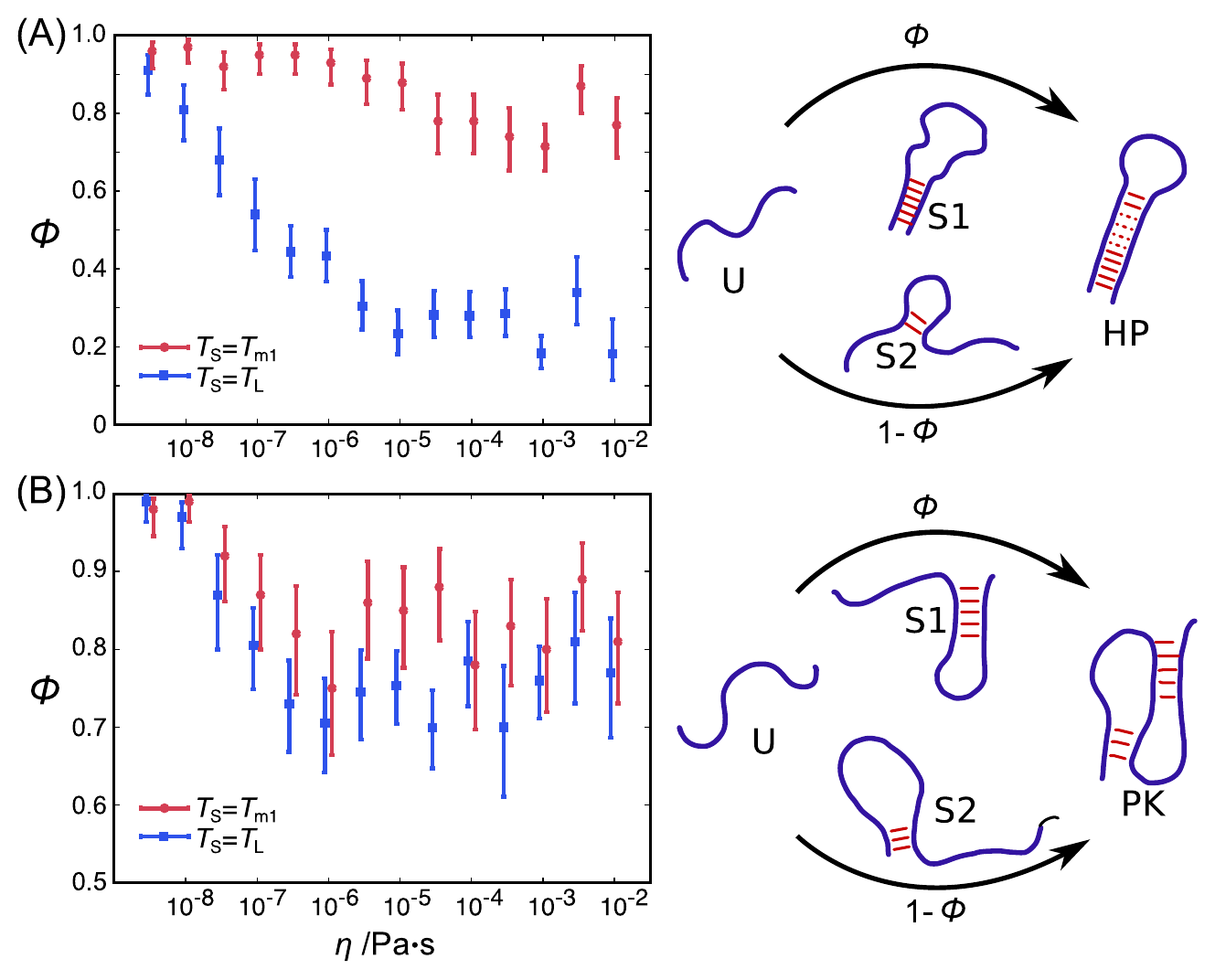}
\caption{\label{fig:Phi} Variations in the flux through the two pathways as
a function of viscosity: (A) hTR HP and (B) BWYV PK. The definition
of states for each RNA is schematically shown on the right.
\textbf{(A)}hTR HP structure naturally splits into two helices, S1 and S2 (Figure~\ref{fig:structure}A). 
The folding pathways are classified if either 
S1 forms first or S2 forms first. The fraction $\Phi$ is the number
of trajectories in which S1 forms first divided by the total number
of trajectories. 
\textbf{(B)} BWYV PK has two hairpin stems, S1 and S2, allowing
us to classify the pathways in the same manner as in panel A. The error
bars indicate 95\% confidence intervals.}
\end{figure}

\paragraph*{\textbf{Viscosity Effects on Hairpin Folding Pathways:}}
The hTR HP has two regions of consecutive canonical base pairs, which
we label stem 1 (S1) and stem 2 (S2) (Figure~\ref{fig:structure}A).
Four noncanonical base pairs are flanked by S1 and S2. Because of
the differences in base pairing between these regions, the folding
pathways may be visualized in terms of formation of S1 and S2 separately. It
is clear that S1 is more stable than S2, and we expect the former
to form first in the folding process according to the stability principle
suggested by Cho, Pincus, and Thirumalai (CPT) \cite{Cho2009}. In
order to assess if the difference in stability leads to friction-induced
changes in the flux between the two pathways (S1 forms before S2 or
\textit{vice versa}), we calculated the fraction of pathways ($\Phi$)
from the folding trajectories, which is obtained by counting the number
of trajectories that reach the folded state by first forming S1. We found
different trends between the two temperatures (Figure~\ref{fig:Phi}).
At $T_\mathrm{m1}$, the dominant pathway (labeled I) is characterized by
formation of the more stable S1 at all values of $\eta$. The flux
through I is $\approx0.8$ at $\eta$ values close to $\eta_{\mathrm{w}}$
(the water viscosity) and that of minor pathway (II) $(1-\Phi)\approx0.2$
in which S2 forms first followed by S1 (figure~\ref{fig:Phi}A). This
finding is in accord with the expectation based on the relative stabilities
of S1 and S2 \cite{Cho2009}. The dominance of pathway I at $T_\mathrm{m1}$
suggests that folding starts away from the loop with the formation
of a base pair between nucleotides G1 and C29 and the HP forms by a
zipping process. 

\onecolumngrid

\twocolumngrid

Interestingly, at the lower temperature $T_\mathrm{L}$, we find that $\Phi$
changes substantially as $\eta$ increases (Figure~\ref{fig:Phi}).
At $\eta$ in the neighborhood of $\eta_{\mathrm{w}}$, $\Phi$ is only $\approx0.2$,
which implies that at $T_\mathrm{L}$ folding predominantly occurs in the
less dominant pathway (II), by first forming the less stable S2. This
finding may be understood using our previous study on P5GA, a 22-nucleotide
RNA hairpin containing only WC base pairs \cite{Hyeon08JACS}. We found
that, although there are multiple ways for P5GA to fold, the most
probable route is through formation of a short loop (SL) that initiates
nucleation of base pair formation involving nucleotides close to the
loop. With that finding in mind, we can rationalize the flux changes
at $T_\mathrm{L}$. The entropy loss ($\Delta S$) due to loop closure, which
in hTR HP would bring the two uracil bases (Figure~\ref{fig:structure})
close enough to initiate a G--C base pair (nucleation step), would be
small ($T\Delta S$$\approx k_{\mathrm{B}}T\ln5$). Once the G--C base
pair near the loop forms, zipping occurs leading to HP formation.
At $T_\mathrm{m1}>T_\mathrm{L}$, S1 formation occurs first, which necessarily involves
long loop (LL) formation that brings $5^{\prime}$ and $3^{\prime}$
ends close. At high temperature this process is facile even though
$T\Delta S\approx k_{\mathrm{B}}T\ln30$. When the $5^{\prime}$ and
$3^{\prime}$ are close, the highly favorable enthalpy gain due to
the formation of a number of favorable WC base pairs compensates for
the entropy loss due LL formation.

\begin{figure}
\includegraphics{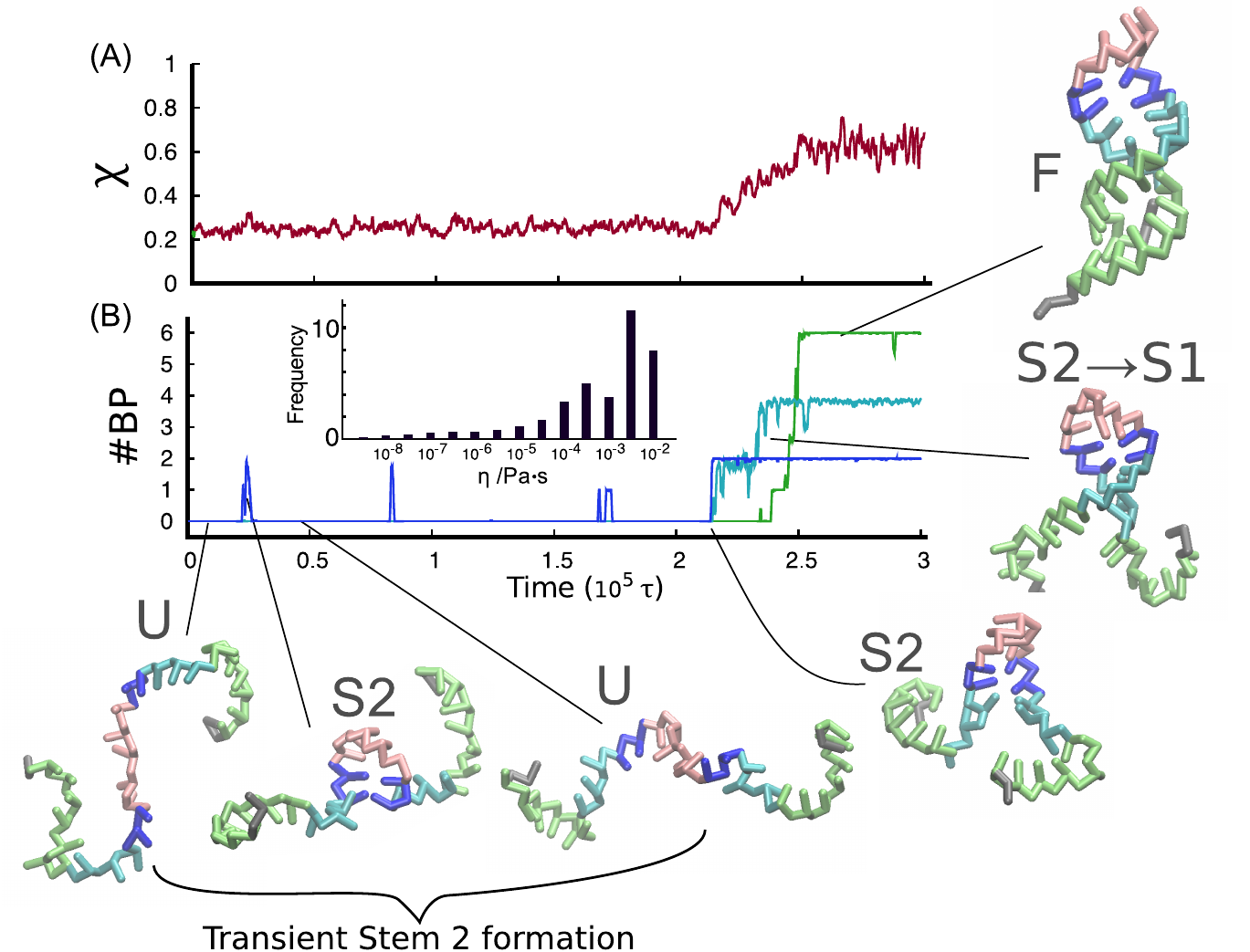}
\caption{\label{fig:traj_HP_LowT} A typical folding trajectory of hTR HP simulated
at high viscosity ($\eta=10^{-3}\,\mathrm{Pa\cdot s}$) with $T_\mathrm{S}=T_\mathrm{L}$.
Time series in panel A shows the structural overlap function, and that in panel B shows the number
of base pairs formed in each stem region 
(blue, stem 2; cyan, noncanonical; green, stem 1) 
along with several snapshots of representative
conformations. In this trajectory, S2 transiently forms three times
before the RNA folds. The folding was initiated with the formation
of S2, followed by the non-canonical base pairs and S1 at last.
\textbf{(B, inset)} Averaged number of the transient S2 formations before hTR
HP reaches the folded state depending on $\eta$.}
\end{figure}

The argument given above to explain
the $\Phi$ values at $T_\mathrm{L}$
can be substantiated by analyzing a typical folding trajectory at
$\eta_{\mathrm{w}}=10^{-3}\,\mathrm{Pa\cdot s}$ shown in Figure~\ref{fig:traj_HP_LowT}.
Before folding occurs, there are several (three times in this particular
trajectory) attempts to form S2 involving the favorable SL, as found
in P5GA hairpin \cite{Hyeon08JACS}. This step is the expected initiation
step in helix nucleation. However, formation of S2, needed for growth
of the helix, is disrupted because S2 is inherently unstable. Consequently,
I2 unfolds and pauses in that state for a long time (Figure~\ref{fig:traj_HP_LowT}).
In the fourth attempt, the formation of two base pairs near the loop
is followed by formation of the noncanonical base pairs, followed
by S1, resulting in the folding of the HP. Interestingly, the transient
S2 formation is only observed at the higher friction regime (Figure~\ref{fig:traj_HP_LowT}B, inset). 
At $\eta>10^{-4}\,\mathrm{Pa\cdot s}$, there are, on average,
5 $\sim$ 10 attempts of S2 formation before the RNA
folds, whereas it does not apparently occur at lower $\eta$, which
is dominated by energy diffusion. 

\onecolumngrid

\twocolumngrid

\paragraph*{\textbf{Viscosity Alters the Flux through the Parallel Routes in BWYV PK Folding:}}
In BWYV PK folding, there are two potential intermediates, I1 characterized
by the formation of the more stable stem 1 or I2 where only stem
2 is formed. In Figure~\ref{fig:Phi}B, we show $\Phi$ as a function
of viscosity. In contrast to the hTR HP case, the pathway through
I1 is always dominant at all values of $\eta$ at both the temperatures.
At the viscosity of water, the fraction $\Phi\approx0.8$. This result
is consistent with experimental studies indicating that I1 is the
major intermediate \cite{Nixon2000,Soto2007}. Our previous study also
showed that the thermal and mechanical (un)folding occur predominantly
through the I1 state \cite{Hori2016}. The present results show that I1 state is
not only thermodynamically stable, but also the major kinetic intermediate. Folding of the PK, which occurs by parallel pathways,
with the dominant one being U$\rightarrow$I1$\rightarrow$F ($\Phi\approx0.8$
at $\eta_{\mathrm{w}}$, for example). In contrast to hTR HP, the loop entropy
in the PK is comparable (Figure~\ref{fig:structure}), and hence the
flux between the two pathways is determined by the CPT stability principle \cite{Cho2009}.

In the dominant pathway, the folding occurs by the following two steps (see Supplemental Movie):
(i) stem 1 folds rapidly after $T$-quench ($\left\langle \tau_{\textrm{U\ensuremath{\rightarrow}I}}\right\rangle =0.02\,\textrm{ms}$
at $\eta=10^{-3}$) forming the intermediate (I1) state, and then
(ii) stem 2 folds after a substantial waiting time ($\left\langle \tau_{\textrm{I\ensuremath{\rightarrow}F}}\right\rangle =0.95\,\textrm{ms}$).
Since there is a large gap in the time scale between the two transitions,
the rate of the whole process ($\textrm{U\ensuremath{\rightarrow}F}$)
is dominated by the second rate determining phase ($\tau_{\textrm{MFPT}}\approx1\,\textrm{ms}$).

\begin{figure}
\includegraphics{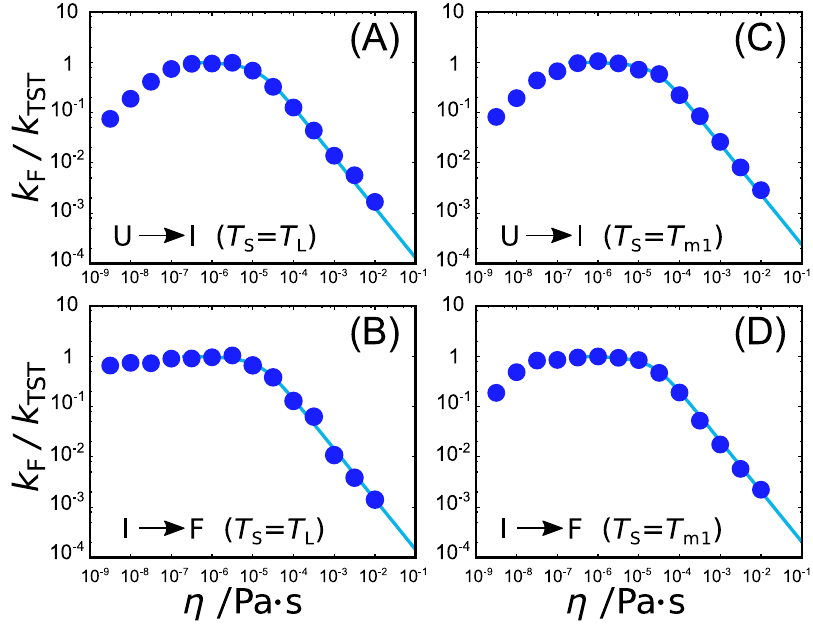}
\caption{\label{fig:kF_2steps}For BWYV PK, folding rates are individually
calculated for two sequential structural transitions through the intermediate,
$\textrm{U}\rightarrow\textrm{I}$ (upper panels) and $\textrm{I}\rightarrow\textrm{F}$
(lower panels). The results of the whole process, $\textrm{U}\rightarrow\textrm{F}$,
are shown in Figure~\ref{fig:kF}C, D.}
\end{figure}

\paragraph*{\textbf{Frictional Effects on Individual Steps in Folding:}}
We have already shown that the rates for the whole folding process ($\textrm{U}\rightarrow\textrm{F}$)
of BWYV PK depend on the viscosity in accord with Kramers' theory
(Figure~\ref{fig:kF}). Since there is a major intermediate, I1, in
the reaction process, we analyzed the folding rates by decomposing
folding into two consecutive reactions, $\textrm{U}\rightarrow\textrm{I}$
and $\textrm{I}\rightarrow\textrm{F}$. Figure~\ref{fig:kF_2steps}
shows the frictional dependence of the folding rates for the two transitions;
$k_{\textrm{I\ensuremath{\rightarrow}F}}$ shows almost the same behavior
as $k_{\textrm{U\ensuremath{\rightarrow}F}}$ since the two time scales
are essentially the same (compare Figure~\ref{fig:kF_2steps} and
Figure~\ref{fig:kF} C, D). It is interesting that the rate of
the faster transition, $k_{\textrm{U\ensuremath{\rightarrow}I}}$,
also exhibits the Kramers-type dependence especially in the high friction
regime, that is, $k_{\textrm{F}}\propto\eta^{-1}$ for $\eta\apprge10^{-5}$
$\textrm{Pa\ensuremath{\cdot}s}$. This result indicates that, even
if the folding reaction involves intermediates, (i) the entire rate
still exhibits the Kramers-type dependence, at least in a case that
one of the substeps is rate limiting, and (ii) a substep that is
not rate determining to the entire rate constant may also show Kramers-type
viscosity dependence.

\section*{Discussion}

\paragraph*{\textbf{Effect of Hydrodynamic Interactions:}}
In order to ensure that our conclusions are robust, we
also examined the effect of hydrodynamic interactions by performing
simulations only at the water viscosity ($\eta_{\mathrm{w}}=10^{-3}\,\textrm{Pa\ensuremath{\cdot}s}$)
for both hTR HP and BWYV PK. As shown in Figure~\ref{fig:kF} (red
circles in B and D), the hydrodynamic interaction (HI) accelerates the
folding rates, but its effect is not as  significant as changing the viscosity.
At $\eta_{\mathrm{w}}$, hTR HP folds with $k_\mathrm{F}\sim9.5\,\mathrm{ms^{-1}}$
with HI, whereas $k_\mathrm{F}\sim6.5\,\mathrm{ms^{-1}}$ without HI. Thus,
the reaction is about 1.5 times faster if HI is included. In BWYV PK case, $k_\mathrm{F}\sim1.9\,\mathrm{ms^{-1}}$
with HI, whereas $k_\mathrm{F}\sim1.1\,\mathrm{ms^{-1}}$ without HI, leading
to a factor of $\sim$1.7 increase, which is similar to the
hTR HP case.

\paragraph*{\textbf{Changes in Viscosity Alter the Flux between Parallel Assembly of RNA:}}
It is well accepted that RNA in general and PK in particular fold
by parallel pathways \cite{Pan97JMB,Cho2009,Roca2018PNAS}. Recently,
it was shown unambiguously that monovalent cations could change
the flux to the folded state between the two pathways in the VPK pseudoknot.
Surprisingly, we find here (see Figure~\ref{fig:Phi}) that $\Phi$ could
be also altered by changing the viscosity for both the HP and the
PK. Although the same prediction was made in the context of protein
folding \cite{Klimov1997}, it is difficult to measure $\eta$ dependence
of $\Phi$ because the secondary structures in proteins are not usually
stable in the absence of tertiary interactions. This is not the case
in RNA. For instance, S1 and S2 are independently stable and hence
their folding could be investigated by excising them from the intact RNA.
Consequently, $\Phi$ as a function of $\eta$ can be measured. Based on the results in Figure~\ref{fig:Phi} showing that
by varying $\eta$ or $\eta$ and $T$, our prediction could be tested
either for the hTR HP or the extensively studied PK (BWYV or VPK).
For example, at $T_\mathrm{L}$ we find that $\Phi$ changes from 0.2 to
0.4 as $\eta$ is varied over a broad range for
hTR HP.   Although
not quite as dramatic, the changes in $\Phi$ are large enough for BWYV PK to be detectable. The
stabilities of the independently folding S1 and S2 constructs can
be also altered by mutations. For instance, by converting
some of the non-canonical base pairs neighboring S2 to WC base pairs
 in the hTR
HP would increase the stability of S2.  Because there are a variety
of ways (concentration of ions, temperature, and mutations) of altering
the independently folding units of RNA, our prediction that $\Phi$
changes with $\eta$ could be readily tested experimentally. 

\paragraph*{\textbf{Speed Limit for RNA Folding:}}
Based on the idea that a protein cannot fold any faster than the fastest
time in which a contact between residues that has the largest probability
of forming, it has been shown that the speed limit ($\tau^{}_{\mathrm{SL}}$)
for protein folding is $\tau^{}_{\mathrm{SL}}\approx1\,\mathrm{\mu s}$ \cite{Kubelka04COSB}.
With the observation that the typical folding barrier height scales
as $\sqrt{N}$ (see Eq.~10 in Ref.~\cite{Thirum95JPI}) and analyses
of experimental data \cite{Li04Polymer}, it was shown that $\tau^{}_{\mathrm{SL}}\approx\tau_{0}\approx(1-10)\mu s$,
where $\tau_{0}=\frac{2\pi\gamma}{\omega_{a}\omega_{b}}$ is the inverse
of the prefactor in Eq.~\ref{eq:kKR_H}. A similar style of analysis
of the experimental data shows that for RNA $\tau^{}_{\mathrm{SL}}\approx1\,\mathrm{\mu s}$ \cite{Hyeon2012BJ}.
Here, an estimate of $\tau^{}_{\mathrm{SL}}\approx\frac{2\pi\gamma}{\omega_{a}\omega_{b}}$
using the values of $\omega_{a}$ and $\omega_{b}$ in Table~\ref{tab:fit}
and $\gamma$ corresponding to water viscosity yields 0.7 $\mathrm{\mu s}$
for the HP and 3.7 $\mathrm{\mu s}$ for the PK. Alternatively, the
value of $\tau^{}_{\mathrm{SL}}=k_{0}^{-1}$ where $k_{0}=k_\mathrm{F}\exp(0.91N^{0.5})$
($k_\mathrm{F}$ is the folding rate obtained using simulations) gives $\tau^{}_{\mathrm{SL}}\approx1$ $\mathrm{\mu s}$
for the HP and $\tau^{}_{\mathrm{SL}}\approx6$ $\mathrm{\mu s}$ for the PK. If $\tau^{}_{\mathrm{SL}}$
is equated with the transition path time, then we can compare estimates
made for DNA hairpins \cite{Truex15PRL} and for RNA constructs (several
PKs and the \textit{add} riboswitch) \cite{Neupane12PRL} obtained
using single molecule experiments. The values range from about 1 to 10 
 $\mathrm{\mu s}$. Thus, there are compelling reasons to assert from
the present and previous theoretical and experimental studies that
an RNA cannot fold any faster than about 1 $\mathrm{\mu s}$.

\paragraph*{\textbf{Influence of Dielectric Friction:}}
In this article, we have treated the electrostatic interactions implicitly,
and hence only systematic and viscous dissipative forces act on the
interaction sites of RNA. We have not considered the effects of dielectric
friction, which could be significant even for an ion moving in an electrolyte
solution \cite{Zwanzig63JCP,Zwanzig70JCP,Hubbard77JCP,Hubbard78JCP}.
In RNA folding, the many body nature of the problem makes it difficult to estimate the magnitude  of the dielectric friction. There are
multiple ions, with significant ion--ion correlations, that condense
onto the RNA in a specific manner dictated by the architecture of
the native fold \cite{Denesyuk2015}. The magnitude of dielectric friction
in this many body system of highly correlated ions could be significant,
which in turn could affect the kinetics of RNA folding. Despite this important
issue, which has not been investigated to our knowledge, it is comforting
to note that experiments as well as simulations reporting viscosity
effects on RNA folding appear to be in accord with Kramers' theory. 

\paragraph*{\textbf{Transmission Coefficients:}}
The ratio $\kappa = \frac{k_\mathrm{F}}{k_{\mathrm{TST}}}$ shown in Figure~\ref{fig:kF} can be as small as $\approx 10^{-3}$, in the high viscosity region. Recently, based on transition path velocity as a measure of recrossing dynamics \cite{Berezhkovskii18JCP}, the values of $\kappa$ have been measured in single molecule pulling experiments \cite{Neupane18PRL} for several DNA hairpins. By fixing the mechanical force at the transition midpoint, where the probability of being folded and unfolded are equal, the folding trajectories were used to estimate that $\kappa \approx 10^{-5}$ \cite{Neupane18PRL}.   
For RNA hairpins, it is known that folding times obtained by $T$-quench are larger by at least 1 order of magnitude relative to times obtained by quenching the force \cite{Hyeon2005}.
Thus, the calculated values of $\kappa$ are not inconsistent with experiments on DNA hairpins under force. It would be most interesting to examine the viscosity dependence of $k_\mathrm{F}$ by maintaining the RNA molecules under tension.

\section*{Conclusions}
Using the TIS coarse-grained model, we investigated the thermodynamics
and folding kinetics of a hairpin and an H-type pseudoknot RNA molecule,
focusing on the dependence of the folding rates on the solvent viscosity.
From temperature-quench folding simulations, we showed that the folding rates
follow the so-called Kramers turnover; the rate increases in the low
friction regime and decreases at high friction, with a maximum rate
at moderate friction. For both the hairpin and the pseudoknot, the
dependence of the folding rates between moderate and high friction
regime is robust and is in accord with the Kramers' theory. We find
clear $\eta^{-1}$ dependence in the folding rates, leaving little
doubt that RNA folding involves a diffusive search in an effective
low dimensional folding landscape. 

A major potentially testable prediction is that in the
$\eta$ values that are accessible in experiments the flux between
pathways by which RNA folds depends on $\eta$. Because the stabilities
of the individual stems could be altered in RNA easily, our prediction is amenable to experimental test.

\section*{Supporting Information}
Movie of a representative trajectory of BWYV PK folding.

\section*{Acknowledgement}
D.T. thanks Bill Eaton for numerous discussions over the years about the role of friction in 
many problems involving biomolecular dynamics.
N.H. is grateful to Debayan Chakraborty, Mauro Mugnai, and Huong Vu for valuable discussions.
We thank the Texas Advanced Computing Center at The University of
Texas at Austin for providing computational resources. 
This work was supported in part by grants from the National Science
Foundation (CHE 16-36424). D.T. also acknowledges additional support
from the Collie-Welch Regents Chair (F-0019) administered through
the Welch Foundation.

\clearpage

\bibliography{friction_Hori}

\end{document}